\documentclass[prb,twocolumn,superscriptaddress,floatfix,noshowpacs,10pt,longbibliography]{revtex4-1}%
\usepackage{graphicx,bm,times}
\usepackage{amsmath}
\usepackage{amsfonts}
\usepackage{amssymb}
\usepackage{bm}
\usepackage{color}
\usepackage{times}
\usepackage{float}

\usepackage{graphicx} 
\usepackage{bm}
\usepackage{color}
\usepackage{times}
\usepackage{xcolor}
\usepackage[%
colorlinks=true,
urlcolor=blue,
linkcolor=blue,
citecolor=blue
]{hyperref}

\begin{document}	
	\title{Strain-tuning of nematicity and superconductivity in single crystals of FeSe}
			
	\author{Michele Ghini}
	\email[corresponding author:]{michele.ghini@studio.unibo.it}
	\affiliation{Clarendon Laboratory, Department of Physics,
			University of Oxford, Parks Road, Oxford OX1 3PU, UK}
	\affiliation{Department of Physics and Astronomy "A. Righi", University of Bologna, via Berti Pichat 6-2, I-40127 Bologna, Italy}
	
	\author{Matthew Bristow}
	\affiliation{Clarendon Laboratory, Department of Physics,
			University of Oxford, Parks Road, Oxford OX1 3PU, UK}

\author{Joseph C. A. Prentice}
	\affiliation{Department of Materials, University of Oxford, Parks Road, Oxford OX1 3PH, United Kingdom}

	\author{Samuel Sutherland}
	\affiliation{Clarendon Laboratory, Department of Physics, University of Oxford, Parks Road, Oxford OX1 3PU, UK}
		
	\author{Samuele Sanna}
	\affiliation{Department of Physics and Astronomy "A. Righi", University of Bologna, via Berti Pichat 6-2, I-40127 Bologna, Italy}

\author{A. A. Haghighirad}
	\affiliation{Clarendon Laboratory, Department of Physics,
	University of Oxford, Parks Road, Oxford OX1 3PU, UK}
	\affiliation{Institute for Quantum Materials and Technologies (IQMT), Karlsruhe Institute of Technology,
76021 Karlsruhe, Germany}

	\author{A. I. Coldea}
	\email[corresponding author:]{amalia.coldea@physics.ox.ac.uk}
	\affiliation{Clarendon Laboratory, Department of Physics,
			University of Oxford, Parks Road, Oxford OX1 3PU, UK}


\begin{abstract}
	
Strain is a powerful experimental tool to explore new electronic states and understand  unconventional superconductivity.
Here, we investigate the effect of uniaxial strain on the nematic and superconducting phase of
 single crystal FeSe using magnetotransport measurements. We find that
 the resistivity response to the strain is strongly temperature dependent and it
  correlates with the sign change in the Hall coefficient being
driven by scattering, coupling with the lattice and multiband phenomena.
Band structure calculations suggest that under strain
the electron pockets develop a large in-plane anisotropy as compared with the hole pocket.
Magnetotransport studies at low temperatures indicate that the mobility of
the dominant carriers increases with tensile strain.
Close to the critical temperature,   all resistivity curves at constant strain
cross in a single point, indicating a universal critical exponent linked to a strain-induced phase transition.
Our results indicate that the superconducting state is enhanced under compressive strain
and suppressed under tensile strain, in agreement with the trends observed in FeSe thin films and overdoped pnictides,
whereas the nematic phase seems to be affected in the opposite way by the uniaxial strain.
 By comparing the enhanced superconductivity under strain of different systems,
our results suggest that strain on its own cannot account for the enhanced high $T_{\rm c}$ superconductivity of FeSe systems.

\end{abstract}

\date{\today}
\maketitle


\section{Introduction}

Uniaxial strain can considerably alter unconventional superconductivity
 \cite{Steppke2017,Malinowski2019,Bohmer2017}
or nematic and magnetic phases in iron-based superconductors \cite{Chu2012,Ikeda2018},
demonstrating that it is a powerful tool to induce phase transitions and explore
the interplay of different competing phases with superconductivity.
A nematic phase is an electronic state of matter in which the electronic
structure develops strong in-plane anisotropy in
transport properties, breaking the rotational symmetry of the  tetragonal lattice.
Strain is also used as a small perturbation to
identify nematic electronic phases
via diverging nematic susceptibility to in-plane anisotropic strain in various iron-based superconductors \cite{Chu2012,Watson2015a,Ishida2018}.
Strain-induced phase transitions can be identified via resistivity scaling nematic critical points \cite{Kuo2016,Hosoi2016}
and  superconductor-insulator quantum phase transitions \cite{Schneider2012,Lin2011,Shi2014}.

FeSe is a unique iron-based superconductor,
which despite its simple structure
hosts a nematic electronic phase, in the absence of a long-range magnetic order.
 This unusual electronic phase is driven by orbitally-dependent effects and correlations that  are responsible for unusual momentum dependent band shifts
  \cite{Baek2015,Tanatar2016,Bohmer2014,Coldea2017}.
  The superconductivity emerging from this nematic phase has a
  two-fold symmetric superconducting gap, orbitally-selective pairing
  \cite{Coldea2017, Sprau2016} and a spin-orbital-intertwined nematic state \cite{Li2020}.
The nematic electronic phase of FeSe is highly sensitive to external parameters,
being strongly suppressed by the isoelectronic substitution with sulphur \cite{Coldea2017,Coldea2020}
and external hydrostatic pressure,  but for higher applied pressure of $\sim 9$~GPa a robust superconducting phase with  a $T_c \sim 40$~K is stabilized, which competes with a spin-density wave \cite{Medvedev2009}.

A monolayer of FeSe, on a suitable substrate, can sustain superconductivity in excess of 65~K, driven by a strong interfacial electron-phonon coupling, the charge transfer through the interface, and strain effects  \cite{Huang2017,Rebec2017}.
 This remarkable superconducting  state is drastically reduced as the number of layers  increases and it is highly dependent on the annealing processes being reduced to close to 20~K  for 50 unit cells  \cite{Wang2017review,Huang2017}.
Superconductivity is enhanced by anisotropic compression in thin films of FeSe on CaF$_2$,
but is suppressed and resistivity increases for films thinner than 100~nm \cite{Nabeshima2013},
 similar to exfoliated flakes of FeSe in the absence of a substrate \cite{Farrar2020}.
Thus, uniaxial strain studies can help to isolate and decouple the different essential components to enhance superconductivity
 and provide important insight in understanding its interplay with the electronic nematic phase of FeSe.

\begin{figure*}[htbp]
	\centering
			\includegraphics[trim={0cm 0cm 0cm 0cm}, width=0.9\linewidth]{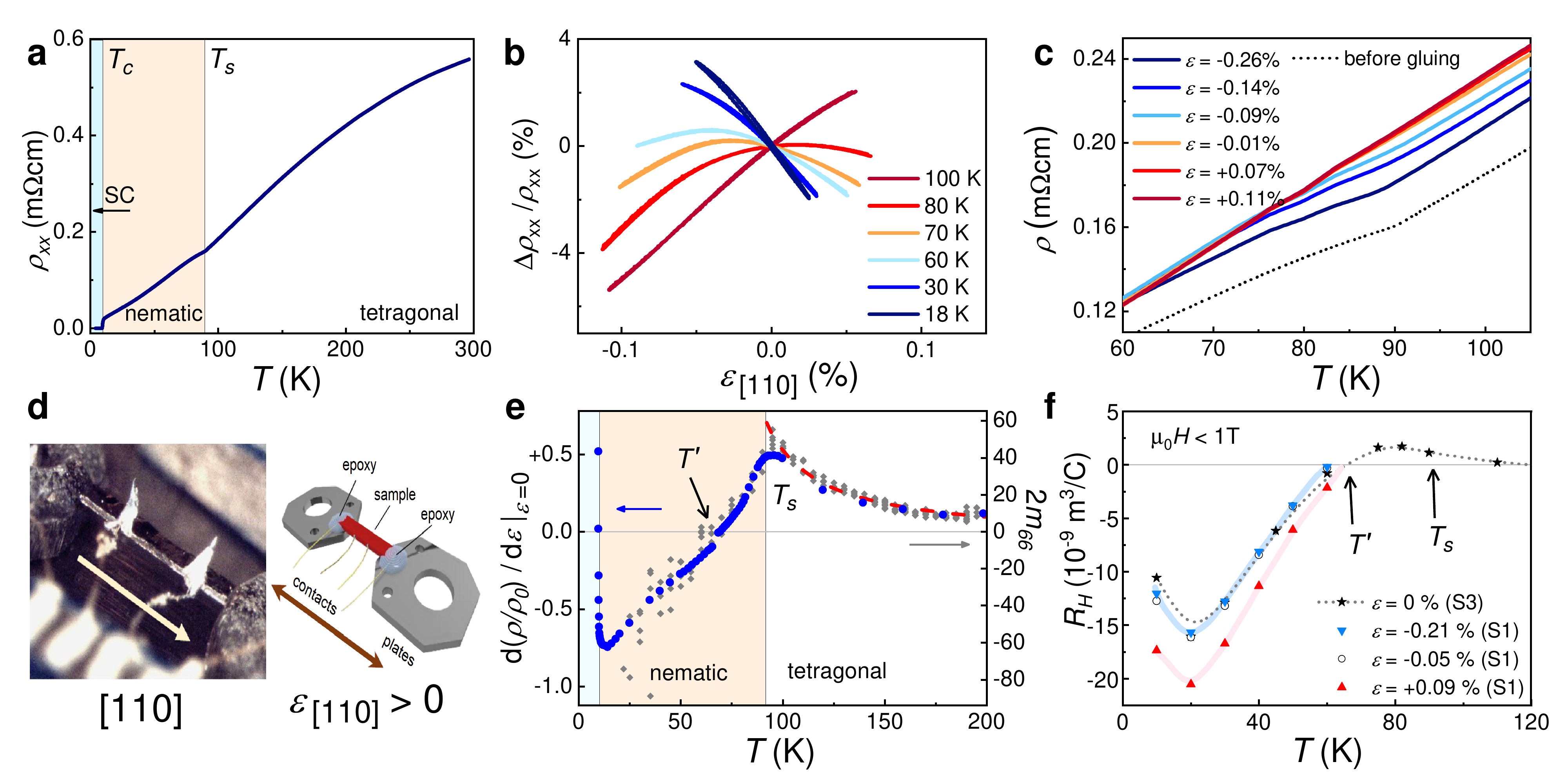}
				\caption{\textbf{(a)} Transport measurement of the single crystal of FeSe with the current along the [110] tetragonal direction.
		\textbf{(b)} Relative changes in resistivity versus applied uniaxial strain at fixed temperatures around the structural transition $T_s$.
		\textbf{(c)} Resistivity measurements as a function of temperature around $T_s$ at different fixed amount of uniaxial strain.
 The dotted line is measured before gluing the sample to the strain cell.
		\textbf{(d)} Single crystal of FeSe glued with epoxy and suspended between the two titanium plates of the strain cell.
		\textbf{(e)} Slope of the linear fit of normalized resistivity against uniaxial strain $d(\rho/\rho_{\varepsilon=0})/d\varepsilon $ for $\varepsilon \rightarrow 0$, indicated by the solid blue symbols. The grey diamonds are the values of $m_{66}$ normalized at 200~K and
the dashed line is a fit to the Curie-Weiss law, after Ref.~\onlinecite{Watson2015a}.
		\textbf{(f)} Hall coefficient $R_H$, estimated  as the low-field slope of the $\rho_{xy}$ versus $B=\mu_0 H$ below 1~T,
measured under different constant uniaxial strain  inside the nematic phase.
Unstrained bulk measurements are indicated by the star symbols, after Ref.~\onlinecite{Watson2015b}.}
\label{fig:F1}
\end{figure*}

In this work, we explore  how the electronic behaviour of bulk single crystals FeSe 
is affected by uniaxial strain, using magneto-transport measurements outside and inside the nematic phase. 
We find that uniaxial strain induces significant changes in resistivity and its gradient is highly temperature dependent being closely correlated with the Hall coefficient inside the nematic phase.
The resistivity curves cross in a single point in the vicinity of the normal to the superconducting transition and we determine a strain-dependent scaling and its critical exponent. Our results under strain in single crystals of FeSe are consistent with those found for epitaxially grown FeSe films on different substrates.
These results indicate that the superconducting state is enhanced under compressive strain and suppressed under tensile strain.

\section{Experimental details}
FeSe single crystal were grown  by the chemical vapour transport method  \cite{Chareev2013,Bohmer2013}.
Electrical connections were made using indium soldering in a 5 point-contact configuration for magneto-transport measurements.
The current flows parallel to the direction of the applied stress,
which is the [110] direction (Fe-Fe bonds) in the tetragonal symmetry and corresponds to the $B_{2g}$ symmetry channel \cite{Willa2019}.
Strain experiments were performed using a CS100 cell from Razorbill \cite{Razorbill}.
The bar shaped single crystal is suspended freely between two mounting plates
 and glued using a two-part epoxy,  different from studies
in which the sample is glued first to a thin titanium plate, which is itself strained
and allows large strain to be applied ~\cite{Park2020}.
 As the material is very soft, the glue itself can also apply a small tensile strain to the sample ($\varepsilon_{\rm glue} \sim  0.02\%$)
 (Fig.~\ref{fig:F1}(d)), as found in previous NMR studies \cite{Baek2016}. 
The capacitance 
between the two plates provides a direct estimate of the applied displacement ($\mu$m).
The amount of nominal stress applied, $\varepsilon = \Delta L/ L_0$  was of the order of $0.05$-$0.25$\%
and the actual lattice distortion was calculated by finite element simulations, as shown in the
 Appendix.

\section{Results and discussion}

\subsection{Resistivity under strain inside the nematic phase}

Fig.~\ref{fig:F1} shows the effect of the applied strain along the [110] tetragonal direction
on the transport behaviour of a single crystal of FeSe.
In the absence of strain, FeSe enters the nematic phase below $T_s \sim $~87~K
and it becomes superconducting at $T_c \sim $~9~K
having a large resistivity ratio between the room temperature resistivity
and that at the superconducting onset  ($RRR \sim 27$)
 (Fig.~\ref{fig:F1}(a)), consistent with previous studies \cite{Watson2015a,Bohmer2016g}.
We performed detailed transport measurements
as a function of uniaxial strain at fixed temperatures,
as the strain cell is capable of applying large and tunable uniaxial stress at low temperatures.
Fig.~\ref{fig:F1}(b) shows the relative variation in resistivity, $\Delta \rho_{xx}/\rho_{xx}(0)=[\rho_{xx}(\varepsilon) -\rho_{xx}(0)]/\rho_{xx}(0)$,
at fixed temperatures inside and outside the nematic phase and in the vicinity of $T_c$.
At high temperatures in the tetragonal phase,
tensile strain (positive strain) increases the resistivity of the system while compressive strain (negative strain) decreases it.
On the other hand, inside the nematic state, the response to strain changes significantly and
 becomes strongly non-linear and the slope changes sign compared with the tetragonal phase.

Fig.~\ref{fig:F1}(e) shows the temperature dependence
of the slope of the normalized resistance as a function of uniaxial strain
 in the limit of small strain ($S=d(\rho/\rho_{\varepsilon=0})/d\varepsilon$, $\varepsilon \rightarrow 0$), extracted from the 
 data shown in Fig.~\ref{fig:F1}(b).
In this low-strain regime, the temperature dependence of the slope follows closely
 the divergent behaviour of the nematic susceptibility, 2$m_{66}$, (measured using piezostacks)
 as approaching $T_s$.  2$m_{66}$ provides a direct measure of the
electronic nematic order parameter and its temperature dependence
 has a Curie-Weiss behaviour, as reported previously in Ref.~\onlinecite{Watson2015a} (see diamond symbols  in Fig.~\ref{fig:F1}e)
 and Refs.~\onlinecite{Tanatar2016,Hosoi2016}.
Inside the nematic phase, the effect of applied strain is unusual and the slope $S$
 changes sign as a function of temperature. We identify a characteristic temperature,  
$T' \sim 70$~K, as the temperature at which the strain has the weakest effect on resistivity and
 the slope changes sign, as compared with the high temperature regime.
 This behaviour is consistent with the sign change of $m_{66}$  below 65~K
 \cite{Watson2015a,Tanatar2016}
as well as the change in resistivity anisotropy induced by the strain of a PEEK substrate \cite{He2018}.
Remarkably, the change in anisotropy at $T'$
coincides also with the temperature
at which  a large anisotropy develops in the local spin susceptibility,
as detected from the line splitting of the Knight shift \cite{Li2020}.
Thus, the changes in the anisotropy of the local magnetism are likely to affect
the scattering and the coherent coupling between local spins and itinerant electrons.

To further address this, we look at
magneto-transport measurements
inside the nematic phase (raw data shown in Fig.~\ref{SMfig:magnetotransport}.
To describe the non-linear effects in
magnetic field of $\rho_{xx}$ and $\rho_{xy}$,
a three-band model was employed to account for a small electron-like pocket,
 besides almost compensated hole and electron pockets \cite{Watson2015b}.
 In the presence of the applied strain, the overall
magnetotransport behaviour does not change significantly and
 still requires a multi-band model to explain these features.
These findings are in contrast to spectroscopic surface-sensitive studies under strain that involve
the presence of a single (uncompensated) electron peanut shape pocket \cite{Watson2017c,Rhodes2020}.
The mobility spectra for bulk FeSe suggest that under tensile strain the mobility of the dominant carriers increase,
as shown in Fig.~\ref{SMfig:magnetotransport}(c) and (d),
in agreement with mobilities of thin films of FeSe under strain \cite{Nabeshima2018}.
Furthermore, the Hall coefficient of bulk FeSe, $R_H=\rho_{xy}/B$ ($B<1$~T) shown in Fig.~\ref{fig:F1}(f),
becomes negative below 65~K under different uniaxial strain, similar to the unstrained case \cite{Watson2015b},
suggesting significant changes in scattering below $T'$.
However, in thin films that have a higher
degree of disorder, the Hall coefficient is always positive as disorder hinders and averages out  the effects responsible for
the negative Hall coefficient \cite{Watson2015b,Nabeshima2018,Bristow2020}.
 An extremum of the Hall coefficient has been assigned
  to a maximum in the scattering anisotropy by spin fluctuations \cite{Breitkreiz2014}
   or driven by the currents renormalized by vertex corrections
dominated by the majority carriers \cite{Fanfarillo2012}.
Coupling with the lattice of these fluctuations may have additional consequences
such as the suppression of superconductivity and of the nematic critical fluctuations \cite{Paul2017,Reiss2020,Labat2017}.

\begin{figure}[htbp]
\centering	
	\includegraphics[trim={0cm 0cm 0cm 0cm}, width=0.75\linewidth]{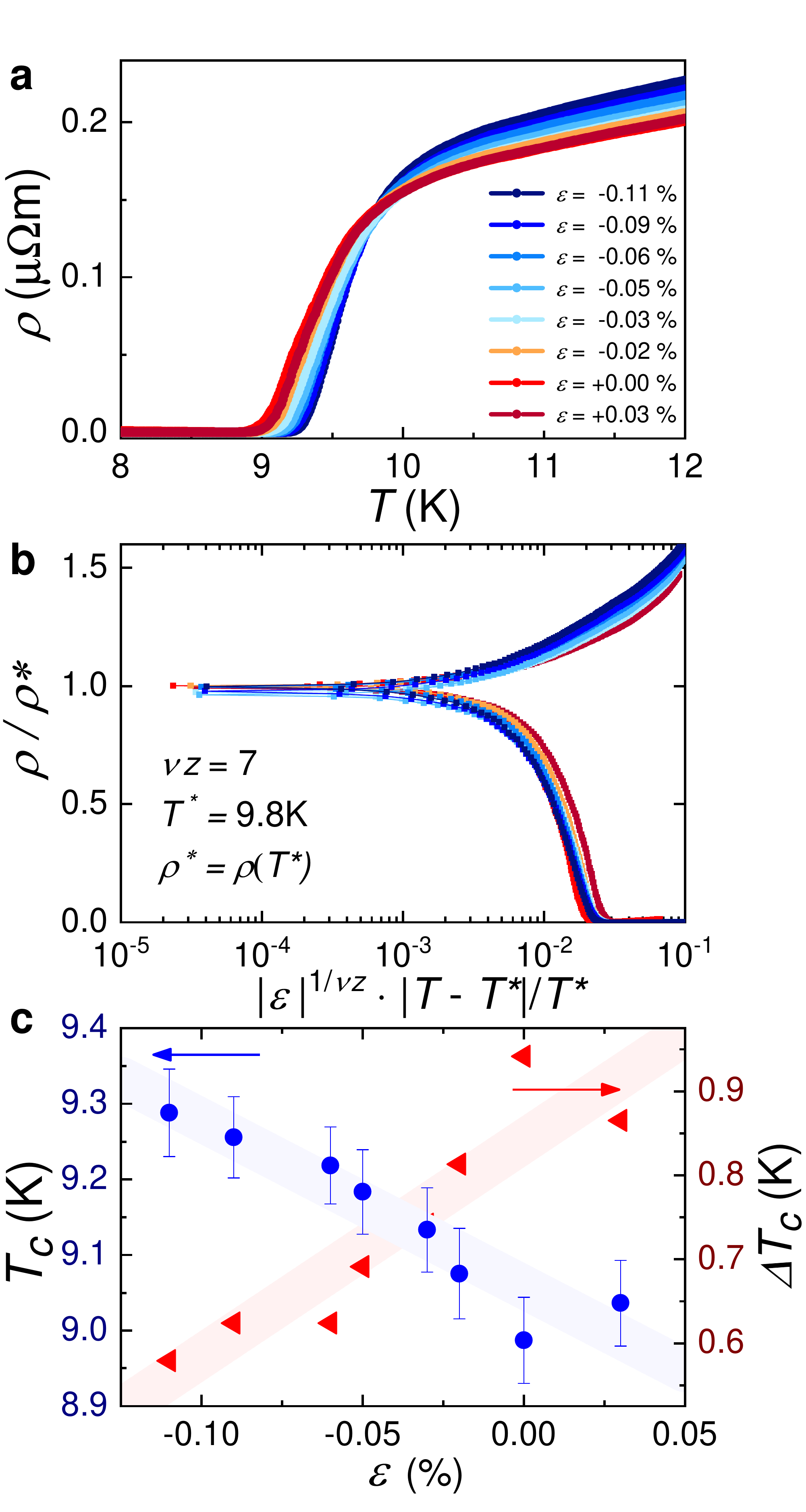}	
	\caption{
		\textbf{(a)} Temperature dependence of the resistivity near the superconducting to normal transition under uniaxial strain.
		 The superconducting state is  clearly enhanced under compressive  uniaxial strain. Opposite trends are found
for tensile strain which can be detected from the constant temperature strain loops in Fig.~\ref{SMfig:scaling_analysis}d.
As the sample is glued to the strain cell, it will be exposed to an additional small tensile strain $\varepsilon_{\rm glue} \sim 0.02 \%$.
	\textbf{(b)} Scaling analysis of superconducting to normal resistivity near the crossing point, $T^* = 9.8$ K, where $\rho^*$ is the resistivity at  $\rho^* = \rho (T^*)$. The ratio between $\rho/\rho^*$ versus $|\varepsilon|^{1/\nu z} \cdot |T-T^*|/T^*$  describes the universality of the transition with a critical exponent of $\nu z \approx 7$.
	\textbf{(c)} Variation of $T_c$, defined as the temperature with zero resistance, and of
the width of the transition, $\Delta T_c$, under applied strain. Solid lines are guides to the eye.	
	}\label{fig:F2}
\end{figure}

To further understand the changes of the electronic structure under strain,
we have calculated the Fermi surface of FeSe.
Fig.~\ref{SMFig:Band_Structure_Calculations}
shows the evolution of the Fermi surface with strain
and for the renormalized and shifted band structure, which was brought in agreement with high temperature ARPES data \cite{Watson2015a}.
 Interestingly, the DFT calculations \cite{Wien2k}  suggest that the in-plane anisotropy is larger for the electron bands
whereas out-of-plane anisotropy  changes for all pockets with increasing strain  (Fig.~\ref{SMFig:Band_Structure_Calculations}).
The sizes and the number of charge carriers for all three pockets shrink with increasing strain suggesting a
potential increase in the Hall coefficient dominated by the most mobile carriers,
as illustrated in Fig.~\ref{fig:F1}(f) and Ref.~\onlinecite{Phan2017ARPES}.
ARPES studies in thin films of FeSe
 indicate that the tensile strain promotes significant shifts of the electron bands
 that can lead to the formation of highly-mobile Dirac carriers \cite{Phan2017ARPES}.
Shifts of 2-3~meV under tensile strain
can reduce the size of both electron and hole pockets \cite{Cai2020};
 the inner electron band could eventually disappear with increasing
 the strength of the orbital order \cite{Coldea2020}.

Next, we focus on the effect of strain on the nematic transition at $T_s$ shown in Fig.~\ref{fig:F1}(c).
In the absence of strain, there is a well-defined anomaly in resistivity at $T_s$,
as shown in Fig.~\ref{fig:SM1}(e).
As the nematic order parameter and associated lattice distortion have a $B_{2g}$  symmetry
that breaks the fourfold rotational symmetry, the applied uniaxial
strain in FeSe along the [110] tetragonal direction induces a finite order parameter at all temperatures.
Therefore, it turns the phase transition into a
crossover,  smearing all the related features at $T_s$
 and the resistivity increases under tensile strain, as the scattering from
nematic domain boundaries becomes significant.
Interestingly, this effect is in the opposite direction to what happens
in the vicinity of the superconducting transition where resistivity
in the normal state is reduced due to tensile strain, as shown in Fig.~\ref{fig:F2}(a).
In our bulk FeSe, the sharp transition at $T_s$
is suppressed from 87~K towards 83~K and replaced by a broad crossover with uniaxial stress 
( in addition to the effect of the glue
that applies $\varepsilon_{\rm glue} \sim  0.02\%$ (Fig.~\ref{fig:SM1}(e,f)).
After applying compressive strain, the resistivity is reduced and the curves seem to recover the signature of the
unstrained sample, as shown in Fig.~\ref{fig:F1}(c).
This behaviour is similar to  that found for  underdoped Co-doped BaFe$_2$As$_2$,
where the feature corresponding to the nematic phase transition  is quickly suppressed (under small strain of $3 \times 10^{-3}$) and replaced by a broad crossover \cite{Ikeda2018}.
In contrast,  the  $\varepsilon_{B1g} $ strain is a continuous tuning parameter 
and has a quadratic variation as a function of strain for
 underdoped Co-doped BaFe$_2$As$_2$ \cite{Ikeda2018}.

\begin{figure}[htbp]
	\centering	
	\includegraphics[trim={0cm 0cm 1cm 0cm}, width=0.8\linewidth]{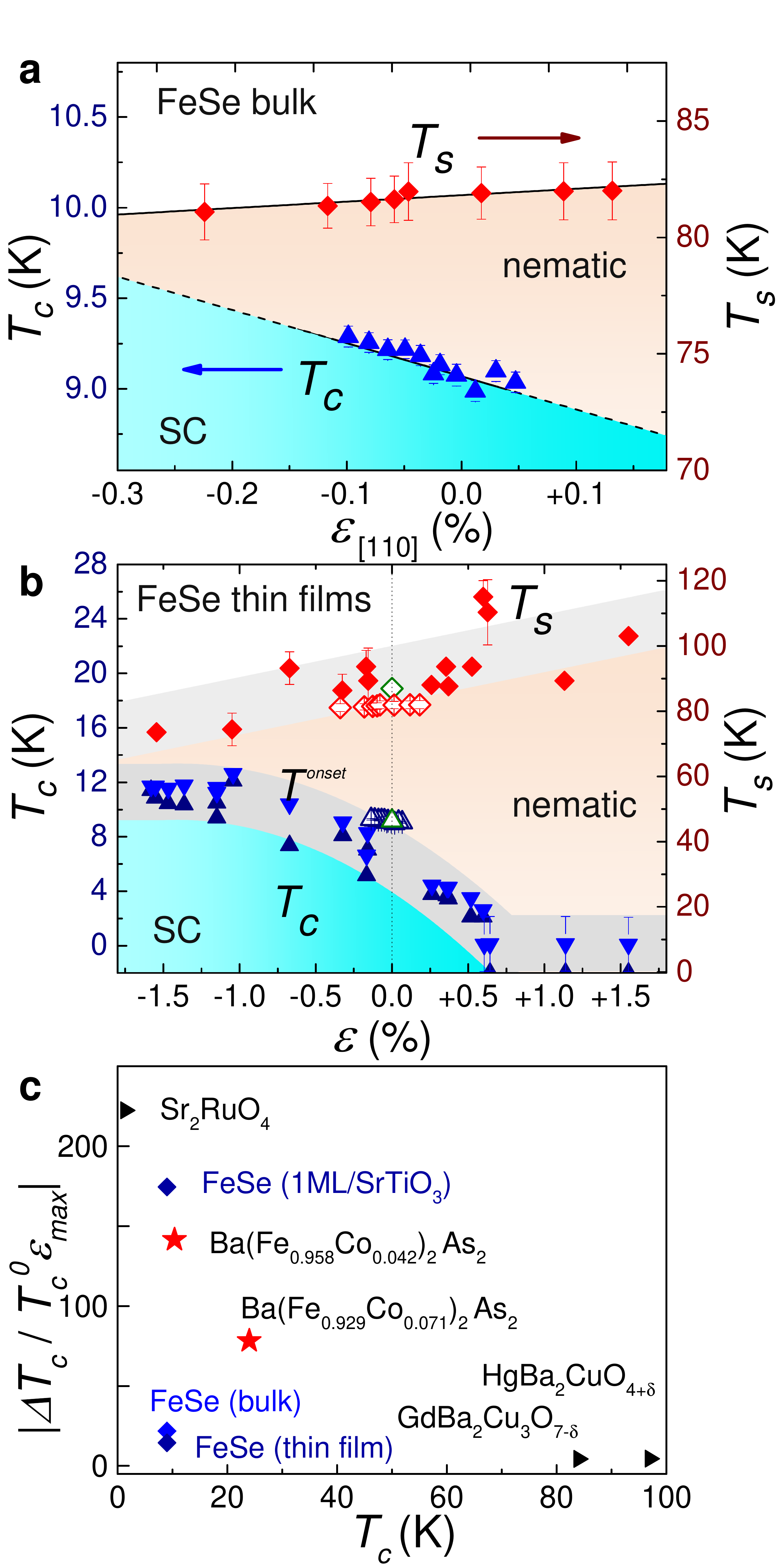}
	\caption{ Superconducting phase diagrams versus strain of \textbf{(a)}  bulk FeSe based on this study
and  \textbf{(b)} thin films of FeSe epitaxially growth on different substrates, adapted after Ref.~\onlinecite{Nabeshima2018}.
Panel \textbf{(b)} contains  the single crystals data (open symbols) from \textbf{(a)}, which are adjusted
to include the additional small tensile strain caused by the glue $\varepsilon_{\rm glue} \sim 0.02 \%$.
 Compressive strain ($\varepsilon < 0$) enhances superconductivity and  seems to suppress the nematic state.
 Black data points in \textbf{(b)} refers to $T_s$ and $T_c$ of bulk FeSe.
\textbf{(c)}  The sensitivity of $T_c$ to strain, $\Delta T_c  $/($ \varepsilon_{max}  \cdot $ $T^{\varepsilon=0}_c$), for a variety of unconventional superconductors,
adapted after Ref.~\cite{Malinowski2019} and compared with bulk, thin films of FeSe and monolayer of FeSe on SrTiO$_3$. }
\label{fig:F3}
\end{figure}

\subsection{Resistivity scaling under strain at low temperatures}

Fig.~\ref{fig:F2}(a) shows the resistivity $\rho(T,\varepsilon)$ of bulk FeSe as a function of the temperature, under various amounts of uniaxial applied strain
 in the vicinity of the superconducting transition. In the normal state the resistivity decreases under tensile strain,
as the mobility of the dominant carriers increases (Fig.~\ref{SMfig:magnetotransport}d).
We find that all resistivity curves, independent from the amount of uniaxial strain applied, cross through a single point in the vicinity of the superconducting transition
around  $T^* \sim 9.8$~K.
Using finite-size scaling analysis,
all data can collapse onto a single curve by re-scaling the resistivity with the value
at the crossing point $\rho^* = \rho$$(T^*)$ and plotting it against the functional relation $|\varepsilon|^{1/\nu z} \cdot |T-T^*|/T^*$, as shown in Fig.~\ref{fig:F2} (b).
Scaling allows the determination of the critical exponents and the universality class of the transition \cite{Vojta2003}.
The best results for bulk FeSe under strain are for a value of $\nu z \approx 7$, as shown in Fig.~\ref{SMfig:scaling_analysis},
larger than $\nu z \approx 7/3 $ found for dirty thin films of FeSe, which describes
the universality class of quantum percolation transitions\cite{Schneider2012}.

Scaling relations to describe superconductor-insulator quantum phase transitions \cite{Schneider2012,Lin2011,Shi2014} and superconductor-metal transitions
report values of $\nu z$ ranging from $0.3$ to $8$ \cite{Xing2015}.
Large values of $\nu z$ can occur when the dynamical critical exponent shows a divergence
by approaching the zero-temperature quantum critical point, as in the case of a Griffiths singularity  \cite{Xing2015}.
A Griffiths phase can be found in the vicinity of a critical point in the presence of disorder for a  two-dimensional superconductor-metal quantum phase transition;
in this case, rare superconducting regions could form in the normal matrix both as a function of temperature and strain.
 Signatures of Griffiths phases have also been detected in the vicinity of a nematic critical point in FeSe$_{0.89}$S$_{0.11}$
 tuned by magnetic field and applied hydrostatic pressure \cite{Reiss2020Griffiths}.
Our findings suggest that strain leads to a superconducting-to-insulating phase transition in bulk FeSe
with the possible formation of inhomogeneous superconducting regions inside the normal matrix.
This important role played by strain and the consequent coupling
with the lattice is likely to have an effect on the critical fluctuations of iron-chalcogenides \cite{Reiss2020}.

\subsection{Phase diagram of FeSe under uniaxial strain}

The resistivity studies of bulk FeSe under strain
have revealed that the compressive strain enhances superconductivity and
narrows the width of the superconducting transition,
as shown in Fig.~\ref{fig:F2}(a) and (c).
Under tensile strain, FeSe displays  a superconductor-to-metal quantum phase transition (see also Fig.~\ref{fig:SM1}d),
similar to underdoped iron-pnictides  \cite{Malinowski2019}.
Fig.~\ref{fig:F3} shows the phase diagrams for bulk FeSe under strain
and thin films of FeSe epitaxially strained by different substrates.
For thin films, the strain is mainly induced by the mismatch in the lattice parameters between the film and the substrate
 \cite{Nabeshima2018,Phan2017}.
 Interestingly, the compressive uniaxial strain
  enhances superconductivity  but it seems to suppress the nematic phase for both bulk and epitaxial films,
  suggesting that the two phases are competing with each other, similar to what is found under small applied hydrostatic and
   chemical pressure \cite{Terashima2016,Coldea2017,Coldea2020}
  However, these opposite trends of the superconductivity versus nematicity under strain
  are in contrast to the effect of growth conditions and disorder \cite{Bohmer2016g}.
  In thin flakes of FeSe,  $T_{\rm c}$  and $T_{\rm s}$ decrease at the same time \cite{Farrar2020} suggesting that
  disorder as well as scattering of twin boundary
  formation inside the nematic phase may have an important role on the transport behaviour. 

The response of  the superconductivity of FeSe to strain can be compared with that
of Co-doped BaFe$_2$As$_2$ single crystals  \cite{Malinowski2019}.
 For underdoped and near-optimally doped compositions,
 the superconducting critical temperature has a quadratic dependence on applied
 strain being rapidly suppressed by both compressive and tensile stress $\varepsilon_{B2g}$.
 In the overdoped regime,  the response of $T_c$ to strain
 is smaller in magnitude and no longer symmetric for
tensile and compressive stress, resembling the behaviour found for FeSe,
which has a nematic phase but no long-range order, similar to overdoped pnictides (Fig.~\ref{fig:F3}(c)).
  The sensitivity to strain of $T_c$ in bulk FeSe agrees with uniaxial high-resolution  thermal-expansion  measurements
  which suggest that superconductivity couples strongly to the in-plane area \cite{Bohmer2013},
 being  anisotropic for the two in-plane directions  ($ dT_c/dp_a$=2.2(5) K/GPa and $dT_c/dp_b$=3.1(1.1) K/GPa  \cite{Bohmer2013})
  whereas under uniaxial strain $dT_c/da$ is $\sim$ 54 K/\AA~  (Fig.~\ref{fig:F3}(c)).
   Thus, to stabilize a 90~K superconductor,  the applied uniaxial strain needs to be extremely large,
   much higher  than the strain generated by a SrTiO$_3$ substrate. This suggests that strain on its own
   is insufficient to enhance superconductivity and  electron doping plays an essential role.

\section{Conclusions.}
To summarize, we have  investigated  the electronic response to
external uniaxial strain of bulk single crystals of FeSe.
We have identified a direct correlation between the transport response to strain, which is
 is strongly temperature dependent, and the temperature changes in the Hall coefficient.
 The normal resistivity increases under tensile strain in the vicinity of
 the nematic transition but it decreases in the proximity of the superconducting transition.
 This suggest that there is a strong coupling between different scattering processes which are
 strongly temperature dependent involving
 the formation of nematic domains, multi-band effects
and the coupling  with the lattice.
 Band structure calculation suggest that under strain
 the electron bands would develop the largest in-plane anisotropy.
Superconductivity of FeSe is enhanced under compressive uniaxial strain, 
similar to overdoped 122 iron-based superconductors, whereas the
the nematic electronic state responds to strain in a opposite way.
Furthermore,  close to the critical temperature, 
we identify a universal crossing point for all resistivity curves measured under constant strain.
The scaling behaviour of  resistivity versus temperature in the vicinity of the superconducting transition  indicates
 the existence of a strain-induced phase transition,
 that would be consistent with the development of rare superconducting regions inside of a normal metal matrix as
 a function of temperature and strain. This study establishes that uniaxial strain,
 on its own, is not sufficient to stabilize a high-$T_c$ FeSe-based superconductor.
 
 {\it Note after review:} After completing this work we become aware of another study of FeSe under strain \cite{Bartlett2021},
in which the single crystals are glued to titanium sheets and these platforms are strained to higher values 
that would be possible for a stand-alone crystal (as in our study). 
The reported findings are in  broad  agreement with the results presented in our work.

 \nonumber
\section{Acknowledgments}

We are very grateful to Dragana Popovic for
helpful communication related to the scaling analysis.
We  thank S. J. Singh and P. Reiss for technical
support, Oliver Humphries for the development of
the mobility analysis spectrum and A. Morfoot for useful comments on the manuscript.
 This work was mainly supported by EPSRC (EP/I004475/1)
and the Oxford Centre for Applied Superconductivity.
M. Ghini acknowledges the financial support of the University of Bologna, Scuola di Scienze. JCAP acknowledges the support of St Edmund Hall, University of Oxford, through the Cooksey Early Career Teaching and Research Fellowship.
The DFT calculations were performed on the University of Oxford Advanced Research Computing Service (https://doi.org/10.5281/zenodo.22558).
AIC acknowledges an EPSRC Career Acceleration Fellowship (EP/I004475/1).\\

\section{Appendix}

\newcommand{\blue}{\textcolor{blue}}
\newcommand{\bdm}[1]{\mbox{\boldmath $#1$}}
\renewcommand{\thefigure}{S\arabic{figure}} 
\renewcommand{\thetable}{S\arabic{table}} 
\newlength{\figwidth}
\figwidth=0.48\textwidth
\setcounter{figure}{0}
\newcommand{\fig}[3]
{	
	\begin{figure}[!tb]
		\vspace*{-0.1cm}
		\[
		\includegraphics[width=\figwidth]{#1}
		\]
		\vskip -0.2cm
		\caption{\label{#2}
			\small#3	
		}
\end{figure}}

\subsection{Strain calibration and simulations}

\begin{figure}[htbp]
	\centering	
	\includegraphics[trim={0cm 3cm 22cm 0cm}, width=0.7\linewidth]{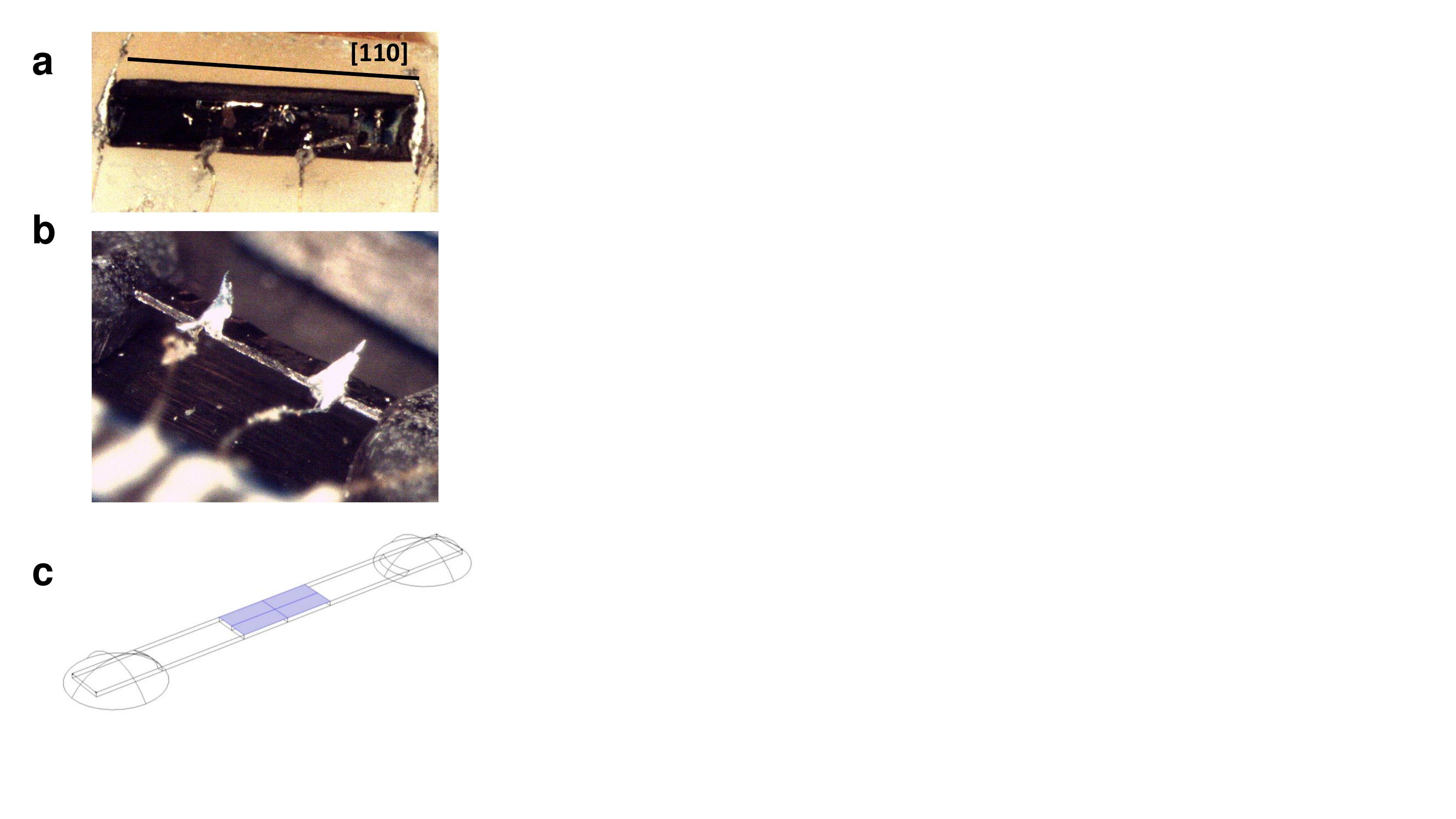}
	\caption{  \textbf{(a)}  Free-standing single crystals of FeSe cut along the [110] tetragonal direction measured before
being mounted on the strain cell in  \textbf{(b)}.
\textbf{(c)} The sample model used to estimate the strain transmission in the sample inside the cell strain secured at both ends by epoxy. }
\label{figSM:sample}
\end{figure}
\begin{figure*}[htbp]
	\centering	
\includegraphics[trim={0cm 0cm 0cm 0cm}, width=0.94\linewidth]{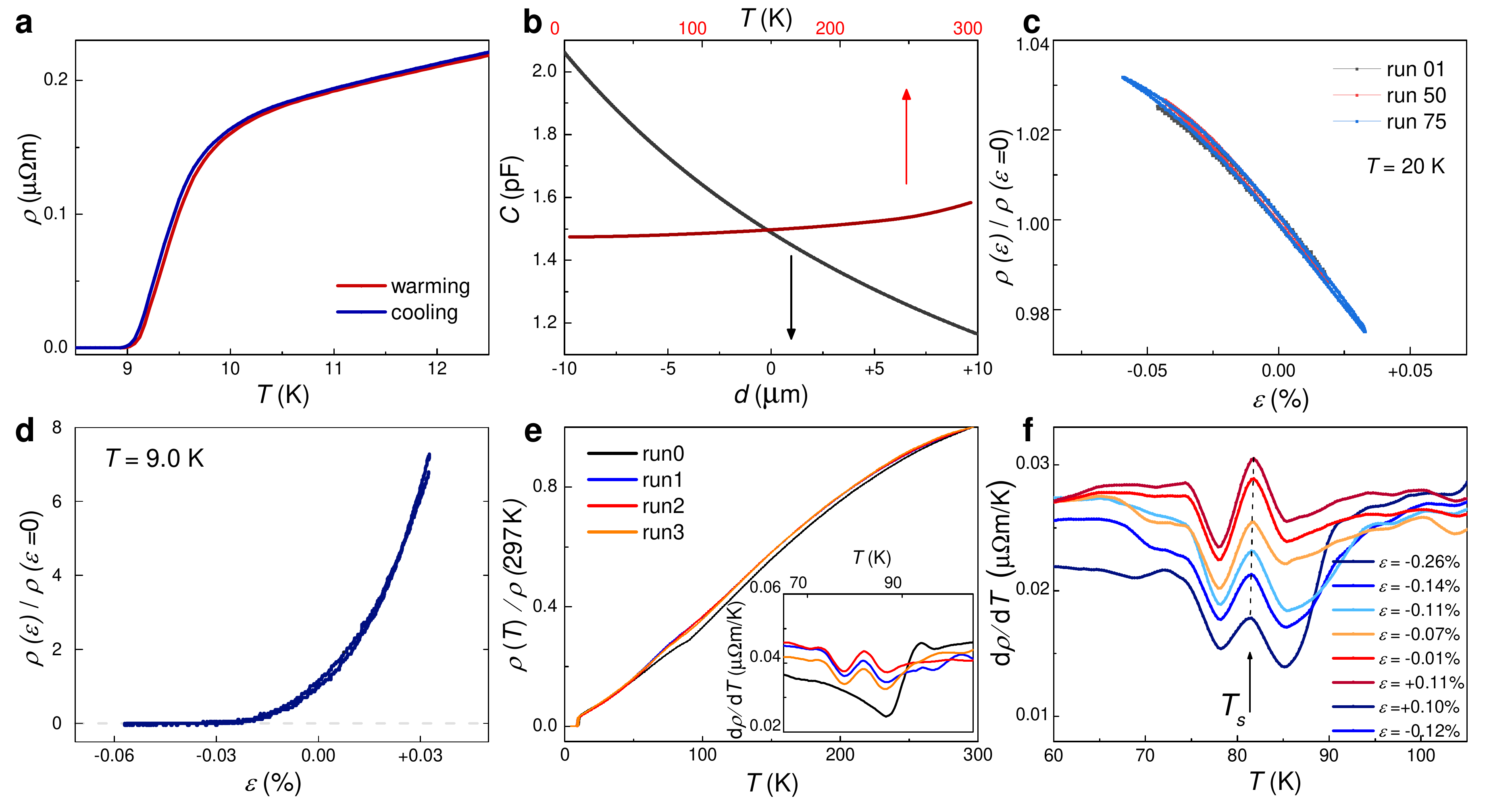}
		\caption{
		\textbf{(a)} Transport measurements performed while cooling and warming,
showing the small temperature variation are due to the thermal load of the strain cell.
		\textbf{(b)} Capacitance as a function of the displacement $d$ ($\mu$m) of the two plates, calculated at room temperature (in black). Temperature dependence of the capacitance without any external strain applied (in red).
		\textbf{(c)} Resistance versus strain at 20~K indicating
the reproducibility and consistency between different measurements acquired at the same temperature.
		\textbf{(d)} Direct evidence that compressive strain enhances the superconducting state
showing the superconducting to normal transition induced by uniaxial compressive strain at fixed temperature $T = 9.0$ K.
		\textbf{(e)} Transport measurements without applied strain at different stages of the experiment.
All the values are normalized at room temperature $\rho$ ($T$) / $\rho$(RT).
Inset shows the derivative of the resistivity against temperature of the same runs, indicating
the smearing of the transition after gluing the sample to the strain cell as compared with the free standing sample (run0).
		\textbf{(f)} Derivative of resistivity against temperature around the nematic transition at different applied strain.
 The structure of the transition developed an additional feature after the gluing of the sample to the strain cell.
 Here, $T_s$ is defined as the position of the Gaussian fit of the middle peak.	}
\label{fig:SM1}
\end{figure*}

Hysteresis strain loops were performed at constant temperature to investigate the sample response to uniaxial tensile and compressive stress.
The hysteresis loops were performed at least twice for each set temperature.
Multiples measurements were conducted on the sample over several weeks to verify reproducibility between different measurements
 (Fig. \ref{fig:SM1} \textbf{(c)}). 

FeSe crystals, with typical dimensions of $\approx 1100\times500\times60$ $\mu$m$^3$, were cut into a rectangular shape along the [110] tetragonal direction, which corresponds applying strain along the $B_{2g}$ symmetry channel  \cite{Willa2019}, and mounted on the CS100 uniaxial strain cell. The amount of nominal stress applied, $\varepsilon_{\rm external} = \Delta L/ L_0$, is defined as the displacement $\Delta L$= $L$($T$)-$L_0$($T)$ divided by the unstrained length $L_0$ of the sample.
An ultra-precise capacitance sensor (Andeen-Hagerling AH270 Capacitance bridge) was used to measure the position of the two titanium plates, in order to evaluate the amount of strain applied with the uniaxial strain cell.
Capacitance as a function of the distance between two plates follows the equation: $ C(d) = {\varepsilon_{\rm external} A}/[d + d_{cell}]$,  with $d_{cell} = 36.00 $ $ \mu$m, $A = 6.04 $ mm$^2$, $  \varepsilon_{\rm external} = \varepsilon_0 \times k = 8.8586 $ pF/m, and where $d$ is the displacement applied from the cell.

In order to asses the  strain transmission through the sample and quantify the internal
strain compared to the applied stress, we have
performed finite element analysis simulations in COMSOL using the Linear Elastic Material material model from the structural mechanics module,
testing our approach against previous reports  \cite{Clifford2014}.
The geometry was constructed and parameterized  to match the geometry of the sample as shown in Fig.~\ref{figSM:sample}.
$\varepsilon_{\rm external}$ was averaged over the the volume between the voltage contacts to get a value for the fraction of strain applied that is transmitted to the sample, and averaged over the top and bottom surfaces between the contacts to get a value for the strain inhomogeneity across the sample.
For the parameters used, $\varepsilon_{\rm [110]}$ = $0.7146$ $\varepsilon_{external}$.
The strain generated by the epoxy  was also simulated and contributes a tensile strain of about $\varepsilon_{\rm glue} \sim  0.02\%$ to the sample.

\begin{figure*}[htbp]
	\centering	
	\includegraphics[trim={0cm 0.5cm 0cm 0cm}, width=0.8\linewidth]{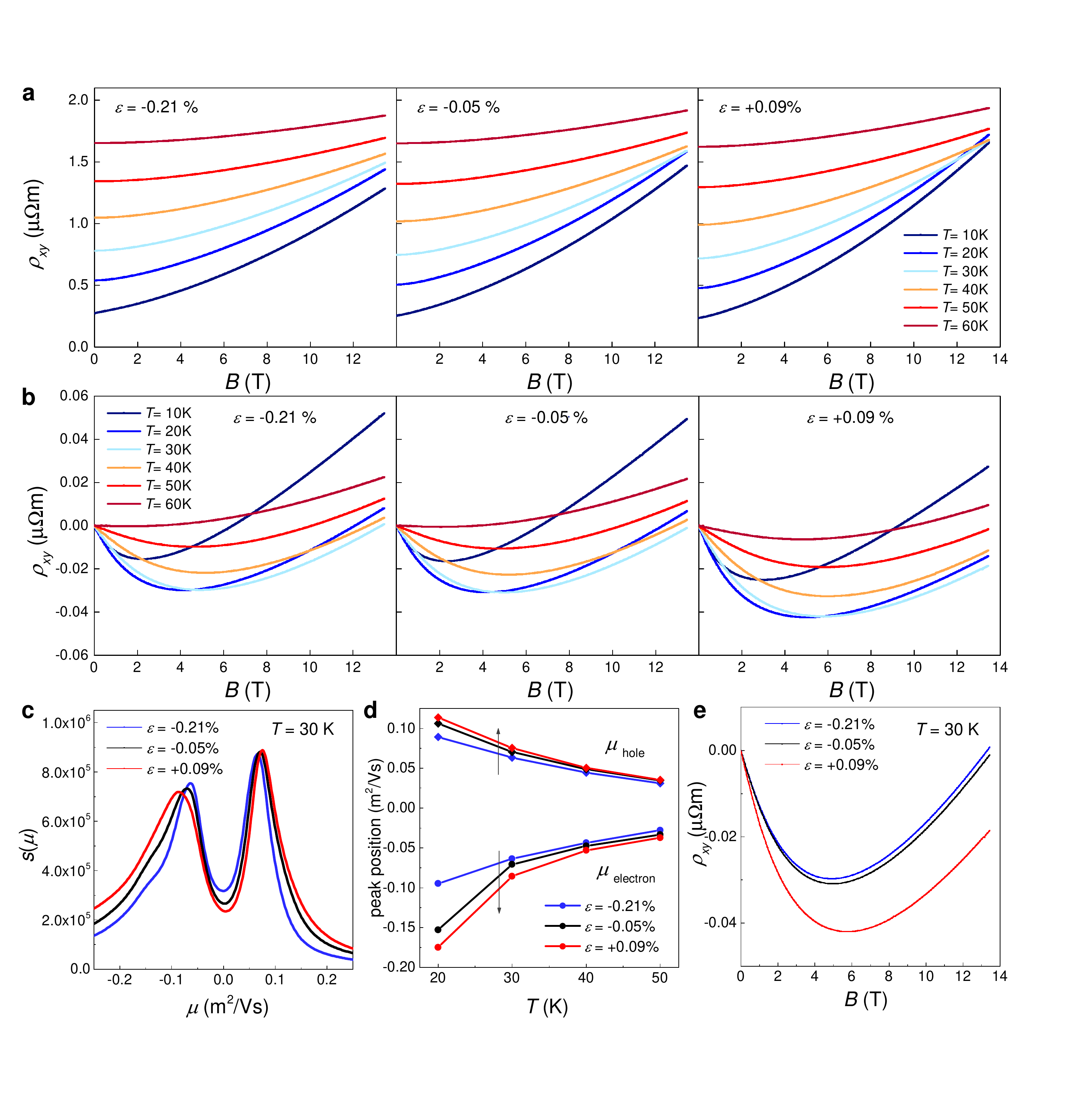}
		\caption{
		\textbf{(a)} The longitudinal resistivity,  $\rho_{xx}$,
and \textbf{(b)} Hall component, $\rho_{xy}$,  measured under different constant uniaxial strain applied along [110] axis for FeSe at
constant temperatures inside the nematic phase.
\textbf{(c)} Mobility spectrum at 30~K extracted from \textbf{(a)} and \textbf{(b)} for different values of uniaxial strain.
 \textbf{(d)} The position of the main peaks from the mobility spectrum for the main electron and hole in the range 20K-50K. For both carriers, tensile strain increases the mobility (in terms of absolute values) while compressive strain tends to decrease it.
\textbf{(e)} Direct comparison between the Hall component $\rho_{xy}$	for different strain at 30~K.}
\label{SMfig:magnetotransport}
\end{figure*}

To extract the value of nominal strain from the measured capacitance, we account for its temperature dependence (the thermal contractions of titanium were negligible). Fig.~\ref{fig:SM1} \textbf{(b)} shows the behaviour of the capacitance against the displacement of the two plates $C(d)$ at room temperature and the dependence of the capacitance sensor versus temperature.
Various measurements were carried out over the temperature range $\Delta T=3-300$ K without any applied strain (zero voltage applied to the piezostacks) in order to establish the trend of $C$($T$,$d_0$).
To account for this effect, the corrected capacitance is obtained by subtracting from the measured capacitance at a given temperature the variation in capacitance between that temperature and room temperature.

As bulk FeSe is a soft layered material, large amounts of strain tend to break the sample. Bending, cracking and exfoliation were observed in various samples with this suspended configuration for higher values of stress.
Fig.~\ref{fig:SM1} \textbf{(e)} shows the resistivity of the sample without applied strain in different stages of the experiment: before and after gluing the sample to the cell (run0 and run1), after all the measurements without magnetic field (run2) and after the last measurement (run3).
The behaviour of the system is consistent between all the different runs (measured over several weeks) but the first one, which corresponds to the free-standing case
which was not glued to the cell.
Fig.~\ref{fig:SM1} \textbf{(f)} illustrates the derivative of resistivity versus temperature in the proximity of the nematic transition, showing the impact of glue on transport properties. It is possible to observe the evolution of $T_s$ under the effect of the applied strain
by assessing the shifts of the high temperature minimum as well as the position of the middle peak.

\begin{figure*} [htbp]
\centering	
	\includegraphics[trim={0cm 0cm 0cm 0cm}, width=0.8\linewidth]{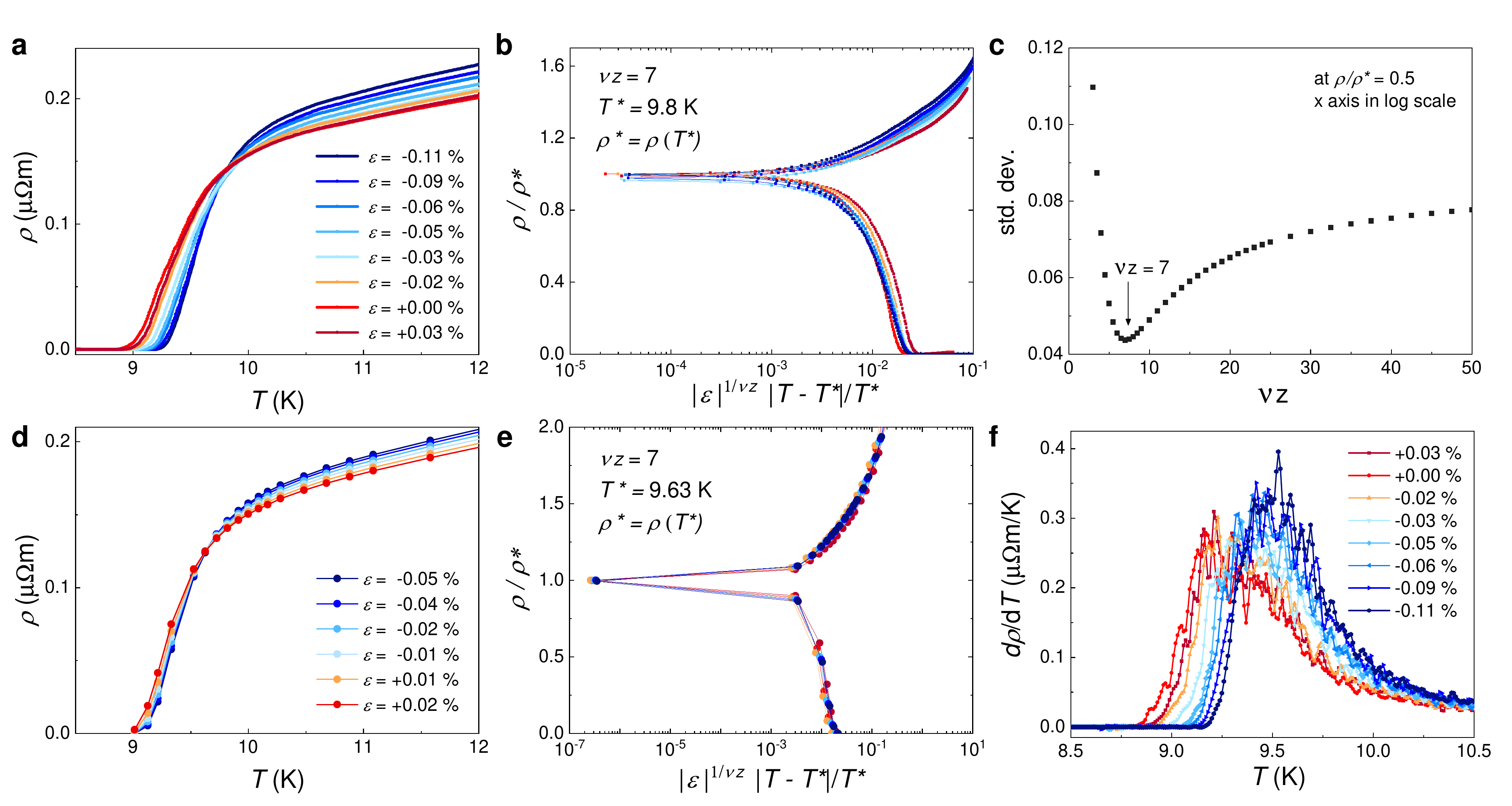}	
	\caption{
		\textbf{(a)} Temperature dependence of the resistivity near the superconductive transition, measurements performed by varying the temperature while keeping constant the applied strain. Negative uniaxial strain (compressive) increases the resistivity while positive uniaxial strain (tensile) has an opposite effect.
The superconductive state is slightly enhanced under compressive strain and suppressed under tensile strain.
		\textbf{(b)} Scaling analysis of (\textbf{a}) near the crossing point at $T^* = 9.8$ K, where $\rho^*$ is the resistivity at the critical point $\rho^* = \rho (T{=}T^*)$. The ratio between $\rho/\rho^*$ vs the functional relation $|\varepsilon|^{1/\nu z} \cdot |T-T^*|/T^*$ is able to describe the universality of the transition with a critical exponent of $\nu z \approx 7$.	
\textbf{(c)} Width of the dispersion of the scaling analysis at $\rho/\rho^* = 0.5 $ as a function of the critical exponent of $\nu z$, for measurements conducted at fixed applied strain (from \textbf{b}).	
		\textbf{(d)} Resistivity near the superconductive transition, with different amount of applied uniaxial strain.
The measurements conducted at constant temperature to confirm findings from previous measurements performed at constant strain.
\textbf{(e)} Scaling analysis of \textbf{d} near the crossing point at $T^* = 9.63$ K, from the data collected at constant temperature
with an exponent of $\nu z \approx 7$, similar to that illustrated in \textbf{b}.
		\textbf{(f)} The first derivative of the data in \textbf{(a)} showing the evolution of width of the superconducting transition with uniaxial strain.	}
\label{SMfig:scaling_analysis}
\end{figure*}

For a limited temperature range in which the capacitance of the cell remains constant, resistivity measurements at constant strain were performed as a function of temperature. Precautions were taken to avoid unwanted effects of thermal drift and fluctuation. All transport measurements as a function of temperature reported in this work were collected during warming ramps and with slow warming rates, in order to limit effects of thermal load at low temperatures. Fig.~\ref{fig:SM1} \textbf{(a)} shows the thermal drift of a cooling ramp compared to the warming measurement.
Fig.~\ref{fig:SM1} \textbf{(d)}
shows the transition between the superconducting to normal state
induced by the uniaxial compressive strain at a fixed temperature ($T = 9.0$~K).

\subsection{Magnetotransport studies}

Magneto-transport measurements were conducted as a function of the magnetic field up to $13.5$ T inside the nematic phase (between $T=10$ K and $T=60$ K), under three different amounts of uniaxial strain between  $\varepsilon = -0.21$\% to $+0.09$\%.
Magneto-transport measurement were conducted measuring both the longitudinal resistivity, $\rho_{xx}$, 
(with the current along the  [110] direction), and the Hall component, $\rho_{xy}$,  as a function of the applied magnetic field applied along the [0 0 1] direction.
These values have been symmetrized and antisymmetrized, with respect to the applied magnetic field (as reported in Fig.~\ref{SMfig:magnetotransport}~\textbf{(a,b)}).
Fig.~\ref{SMfig:magnetotransport}~\textbf{(c)} reports the mobility spectrum generated from the magneto-transport data for different amount of strain at $T=30$~K.
 The mobilities of the dominant charge carriers are consistently enhanced under tensile strain and suppressed under compressive strain,
inside the nematic phase (Fig.~\ref{SMfig:magnetotransport}~\textbf{(d)}).
The low field Hall coefficient, $R_{\rm H}$ (in $B<1$T),  is surprisingly insensitive to compressive strain but changes under tensile strain,
as shown in Fig.~\ref{SMfig:magnetotransport}~\textbf{(e)}.

\subsection{Scaling analysis}

Fig.~\ref{SMfig:scaling_analysis} reports scaling of the reduced resistivity $\rho / \rho^*$ around the strain-independent crossing point $T^*$ as a function of $|\varepsilon|^{1/\nu z} \cdot |T-T^*|/T^*$ (Fig. \ref{SMfig:scaling_analysis} \textbf{(b)}).
This functional relation leads to a collapse of the experimental data with the same exponent $\nu z \approx 7$.
Fig.~\ref{SMfig:scaling_analysis} \textbf{(c)} reports the standard deviation of the dispersion at  $\rho/\rho^* = 0.5 $ versus different values of the critical exponent $\nu z$ as a way to estimate the optimal exponent. Optimal results were obtained with a value close to the minimum $\nu z = 7$.
 Fig.~\ref{SMfig:scaling_analysis} \textbf{(d)} shows transport measurements at the superconductive transition collected by varying the uniaxial strain at fixed temperatures, instead of varying the temperature with fixed strain. As before, we observe that compressive strain increases the resistivity above the transition and enhances the superconductive state (while tensile strain induces opposite effects) and the presence of the  temperature $T^*$ at the middle of the transition.

\subsection{Band structure calculations}

\begin{figure*}[htbp]
\centering	
	\includegraphics[trim={0cm 0cm 1cm 0cm}, width=0.98\linewidth]{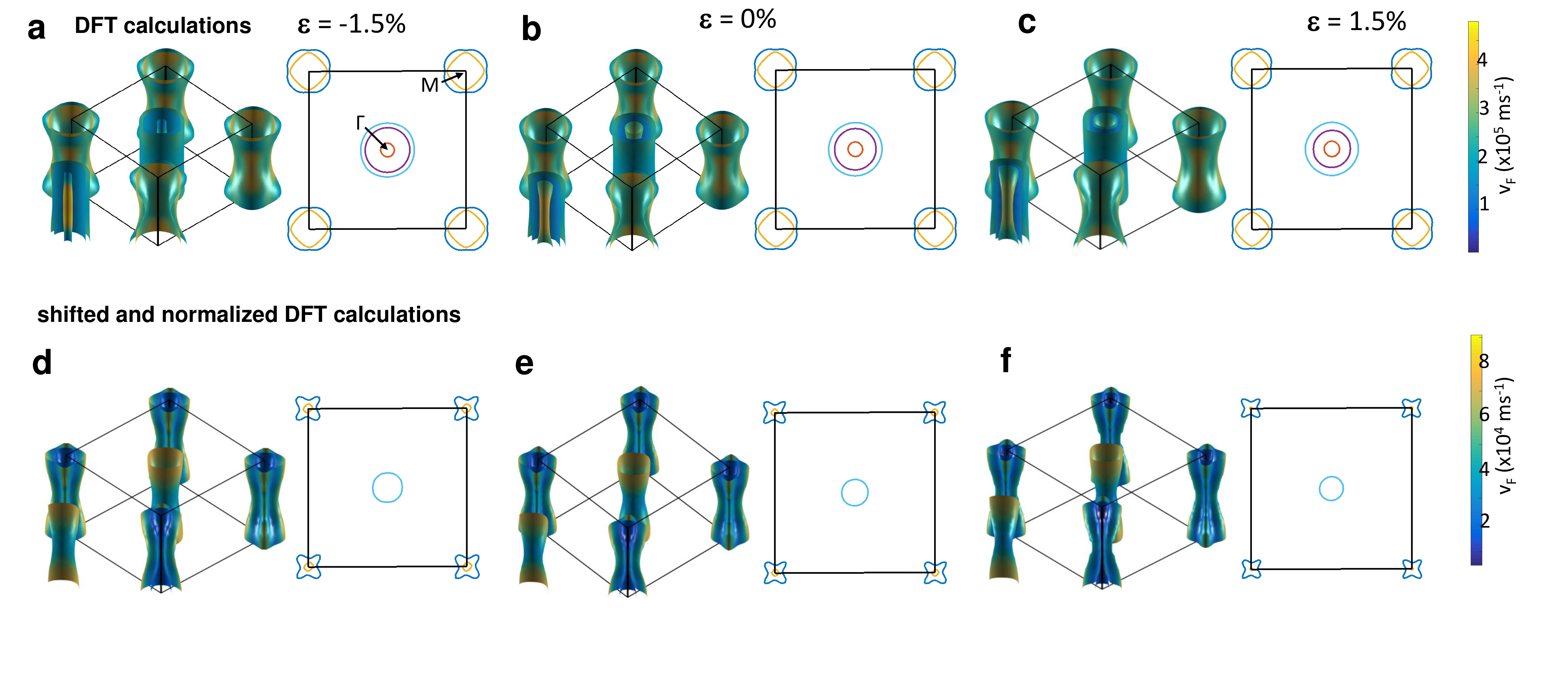}	
	\includegraphics[trim={0cm 10cm 1cm 0cm}, width=0.98\linewidth]{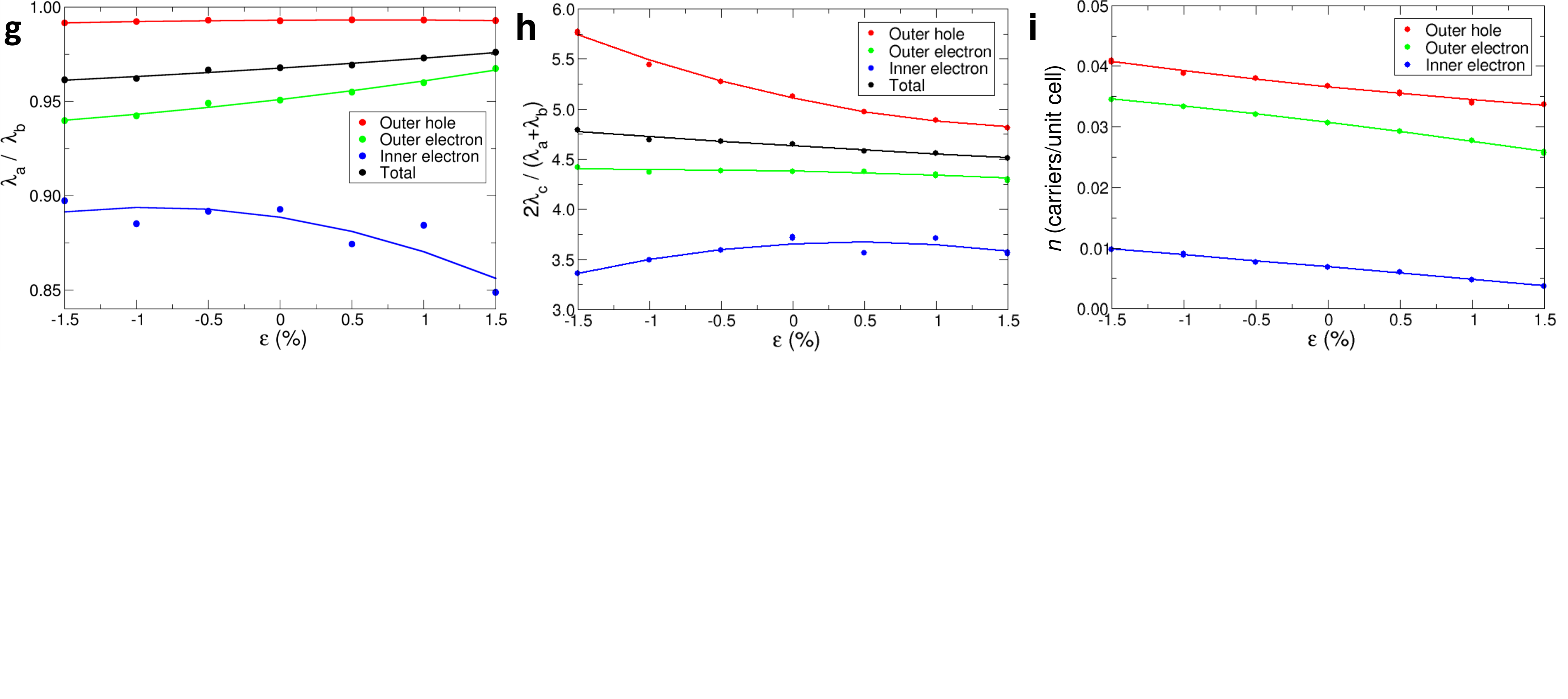}	
    \caption{Calculated Fermi surface plots and slices through the centre of the Brillouin zone ($\Gamma$-M plane) for FeSe
   for different applied strain $\varepsilon$ along [110] of $\varepsilon=-1.5\%$ in  (a, d), $\varepsilon=0\%$ in  (b, e), and $\varepsilon=1.5\%$ in  (c, f).
    3D Fermi surface plots are coloured by the magnitude of the Fermi velocity. Colour bars for each row are displayed on the far right.
    Panels 	\textbf{(a-c)} show the Fermi surface with increasing strain as calculated in Wien2K using
    the GGA approximation (PBE functional) \cite{Wien2k,GGA} including the spin-orbit coupling. We use the
    experimental orthorhombic parameters ($Cmma$ symmetry and $a$ = 5.308~\AA~, $b$ = 5.334~\AA, ~$c$ = 5.486 \AA~
    and the volume cell of the unit cell is conserved under applied uniaxial strain.  Panels 	\textbf{(d-f)}  show the Fermi surface after band renormalization and shifting are applied to match the unstrained experimental ARPES data at high temperatures above $T_s$ \cite{Watson2015a}.  Band renormalization of 3 and 4 are applied to the outer hole band at $\Gamma$ and both electron bands at M respectively,
    and a band shift of 45~meV is applied to the electron bands and  -45~meV is applied to the hole band to ensure charge compensation, as suggested by previous experimental ARPES data \cite{Watson2015a}.
 \textbf{(g)} In-plane and \textbf{(h)} out-of-plane anisotropy of the penetration depth.
 \textbf{(i)} The number of charge carriers per unit cell for each quasi-two-dimensional cylinder of each shifted Fermi surface pocket as a function of strain.
 The solid lines are guides to the eye.}
\label{SMFig:Band_Structure_Calculations}
\end{figure*}

Fig.~\ref{SMFig:Band_Structure_Calculations} (a-c) reports Fermi surfaces and slices through the centre of the Brillouin zone ($\Gamma$-M plane) for bulk FeSe for different applied strain along [110] of $\varepsilon=\pm1.5\%$.
The same renormalizations and band shifts were chosen to agree with ARPES data in the unstrained tetragonal case\cite{Watson2015a}
and these shift were applied at all calculations under strain (Fig.~\ref{SMFig:Band_Structure_Calculations} (d-f)).
The outer hole band and the two electron bands were renormalized by a factor of 3 and 4 respectively, as suggested by ARPES data [6], and then shifted by -45meV and 45 meV respectively, to ensure charge compensation. Additionally, the inner and middle hole bands are shifted away from the Fermi level entirely, as compared with the unshifted case (Fig.~\ref{SMFig:Band_Structure_Calculations} (a-c)).
We find that both the outer hole band and the electron bands shrink  with increasing tensile strain in both the unshifted and shifted cases. This reflects clearly
in the number of charge carriers that decreases with increasing tensile strain for all three Fermi surface pockets,
as shown in Fig.~\ref{SMFig:Band_Structure_Calculations}(i).

The understand the effect of the strain on the in-plane and out-of-plane anisotropy of the Fermi surface pockets b, we have also computed the penetration depth, $\lambda$, along the three lattice vectors, and taking suitable ratios. Fig.~\ref{SMFig:Band_Structure_Calculations}(g) shows the in-plane anisotropy ($\lambda_a/\lambda_b$) and Fig.~\ref{SMFig:Band_Structure_Calculations} (h) shows the out-of-plane anisotropy ($2\lambda_c$/($\lambda_a+\lambda_b$)), calculated from the renormalized and shifted Fermi surface pockets for several values of strain.  The inner electron pocket is much smaller than the others, so numerical errors affect the results much more strongly, as seen. We find that the hole pocket is essentially isotropic in-plane, whilst the electron pockets have some in-plane anisotropy. However, all three pockets exhibit out-of-plane anisotropy, with the hole band exhibiting the largest anisotropy. The anisotropy in the overall penetration depth, calculated by combining the contributions from each pocket, remains fairly constant with strain, both in-plane and out-of-plane.

\bibliography{FeSe_strain_bib_dec20}

\end{document}